\documentclass[12pt]{article}
\usepackage{a4}
\usepackage{amsfonts}
\usepackage{amsmath,amssymb}

\makeatother

\def\un#1{\relax\ifmmode\@@underline#1\else
        $\@@underline{\hbox{#1}}$\relax\fi}

\let\du=\du                     

\def\a{\alpha}
\def\b{\beta}

\def\d{\delta}

\def\f{\phi}
\def\g{\gamma}
\def\h{\eta}

\def\j{\psi}
\def\k{\kappa}

\def\m{\mu}
\def\n{\nu}

\def\p{\pi}

\def\s{\sigma}

\def\D{\Delta}

\def\L{\Lambda}

\def\ve{\varepsilon}

\def\ce{{\cal E}}

\def\cg{{\cal G}}

\def\car{{\cal R}}

\def\cw{{\cal W}}

\def\bo{{\raise-.3ex\hbox{\large$\Box$}}}               
\def\pa{\partial}                                       
\def\de{\nabla}                                         
\def\TH{{\raise.2ex\hbox{$\displaystyle \bigodot$}\mskip-4.7mu \llap H \;}}
\def\face{{\raise.2ex\hbox{$\displaystyle \bigodot$}\mskip-2.2mu \llap {$\ddot
        \smile$}}}                                      

\def\abs#1{\left| #1\right|}                    
\def\leftrightarrowfill{$\mathsurround=0pt \mathord\leftarrow \mkern-6mu
        \cleaders\hbox{$\mkern-2mu \mathord- \mkern-2mu$}\hfill
        \mkern-6mu \mathord\rightarrow$}
\def\dvec#1{\vbox{\ialign{##\crcr
        \leftrightarrowfill\crcr\noalign{\kern-1pt\nointerlineskip}
        $\hfil\displaystyle{#1}\hfil$\crcr}}}           
\def\dt#1{{\buildrel {\hbox{\LARGE .}} \over {#1}}}     

\def\frac#1#2{{\textstyle{#1\over\vphantom2\smash{\raise.20ex
        \hbox{$\scriptstyle{#2}$}}}}}                   
\def\sfrac#1#2{{\vphantom1\smash{\lower.5ex\hbox{\small$#1$}}\over
        \vphantom1\smash{\raise.4ex\hbox{\small$#2$}}}} 
\def\bfrac#1#2{{\vphantom1\smash{\lower.5ex\hbox{$#1$}}\over
        \vphantom1\smash{\raise.3ex\hbox{$#2$}}}}       
\def\afrac#1#2{{\vphantom1\smash{\lower.5ex\hbox{$#1$}}\over#2}}    

\def\[{\lfloor{\hskip 0.35pt}\!\!\!\lceil}
\def\]{\rfloor{\hskip 0.35pt}\!\!\!\rceil}
\def\Lag{{\cal L}}
\def\du#1#2{_{#1}{}^{#2}}

\def\un{\underline}
\def\fracmm#1#2{{{#1}\over{#2}}}

\def\low#1{{\raise -3pt\hbox{${\hskip 0.75pt}\!_{#1}$}}}

\def\Dot#1{\buildrel{_{_{\hskip 0.01in}\bullet}}\over{#1}}
\def\dt#1{\Dot{#1}}

\newskip\humongous \humongous=0pt plus 1000pt minus 1000pt
\def\caja{\mathsurround=0pt}
\def\eqalign#1{\,\vcenter{\openup2\jot \caja
        \ialign{\strut \hfil$\displaystyle{##}$&$
        \displaystyle{{}##}$\hfil\crcr#1\crcr}}\,}
\newif\ifdtup

\newcommand{\be}{\begin{equation}}
\newcommand{\ee}{\end{equation}}
\newcommand{\nbe}{\begin{equation*}}
\newcommand{\nee}{\end{equation*}}

\newcommand{\lb}{\label}

\begin{document}

\thispagestyle{empty}

{\hbox to\hsize{
\vbox{\noindent July 2010 \hfill version 3}}}

\noindent
\vskip2.0cm
\begin{center}

{\large\bf CHAOTIC INFLATION IN F(R) SUPERGRAVITY~\footnote{Supported 
in part by the Japanese Society for Promotion of Science (JSPS) and CERN}}
\vglue.3in

               Sergei V. Ketov~${}^{a,b}$
\vglue.1in

${}^a$ {\it Department of Physics, Tokyo Metropolitan University, Japan}\\
${}^b$ {\it IPMU, University of Tokyo, Japan}
\vglue.1in
ketov@phys.metro-u.ac.jp
\end{center}

\vglue.3in

\begin{center}
{\Large\bf Abstract}
\end{center}
\vglue.1in

\noindent The bosonic $f(R)$ gravity function is derived from a chiral $F(R)$ 
supergravity model for the first time. We find the existence of the upper limit
(or AdS bound) on the scalar curvature, as well as a solution with the 
vanishing cosmological constant. We compare our simple model of $F(R)$ 
supergravity to the well known Starobinsky model of chaotic inflation.

\newpage

\section{Introduction}

Revealing the identity of inflaton and unification of cosmological inflation 
with High-Energy Physics remain the outstanding problems beyond the Standard 
Model of elementary particles and Einstein gravity. One of the easy ways of
realization of an inflationary universe is provided by the popular theories of 
$f(R)$ gravity, whose Lagrangian is a function $f(R)$ of the scalar curvature 
$R$ in four space-time dimensions (see eg., ref.~{\cite{fr} for some recent
reviews). The use of those theories in inflationary cosmology was pioneered by
Starobinsky \cite{star}.

Any $f(R)$ theory of gravity is known to be equivalent to a scalar-tensor 
theory of gravity \cite{eq}. In view of that equivalence, a dynamics of the
spin-2 part of a metric (i.e. gravity itself) is not modified at all, but there
 is the extra propagating scalar field given by the conformal mode of the 
metric. The latter plays the role of inflaton in the inflationary models based 
on $f(R)$ gravity. Being unrelated to any fundamental theory of gravity, those
inflationary models are truly phenomenological and have no connection to
High-Energy Physics. Moreover, there is no mechanism  inside the $f(R)$ gravity
 theories, that would protect a particular choice of the function $f(R)$ 
against quantum corrections that may destabilize inflation or exclude its
slow roll.

In our recent papers \cite{our} we constructed the new 
supergravity theory that can be considered as the $N=1$ locally supersymmetric 
extension of the $f(R)$ gravity.~\footnote{Another (unimodular) $F(R)$ 
supergravity theory was proposed in ref.~\cite{nish}.} Supergravity is 
well-motivated in High-Energy Physics Theory beyond the Standard Model of 
elementary particles. Supergravity is also the low-energy effective action of 
Superstrings.~\footnote{Some applications of $F(R)$ supergravity to Loop
Quantum Gravity were given in ref.~\cite{gky}.} Unlike the $f(R)$ theories
 of gravity, the $F(R)$ supergravity is highly constrained by local 
supersymmetry and consistency. Moreover, our superspace construction 
of $F(R)$ supergravity \cite{our} leads to a {\it chiral\/} action in 
curved $N=1$ superspace, which may be naturally stable against quantum 
corrections that are usually given by {\it full\/} superspace integrals. Our 
supersymmetric extension of $f(R)$ gravity is non-trivial because the 
supergravity auxiliary fields do not propagate (this feature is called 
the auxiliary freedom \cite{gat}). However, the superconformal mode of the
supergravity supervielbein becomes dynamical in $F(R)$ supergravity. As was 
proven in ref.~\cite{our}, an $F(R)$ supergravity is equivalent to the standard
 $N=1$ Poincar\'e supergravity coupled to the dynamical chiral superfield 
whose K\"ahler potential and superpotential are dictated by a single 
holomorphic function. That chiral superfield is precisely the superconformal 
mode of the supervielbein. It was argued in ref.~\cite{our} that the leading
field component of the chiral superfield may be identified with the 
dilaton-axion field in Superstring Theory.
 
The component structure of $F(R)$ supergravity is very complicated, and 
some of its general features were outlined in ref.~\cite{our}. However, 
no explicit derivation of a bosonic (real) function $f(R)$ out 
of the supergravity (holomorphic) function $F(R)$ was given. In this Letter we 
fill out this gap, by giving the first explicit example of such calculation.

In sec.~2 we briefly review our superspace construction of $F(R)$ supergravity
along the lines of ref.~\cite{our}, and formulate the equation for the
 auxiliary fields. In sec.~3 we propose the simplest non-trivial model of 
$F(R)$ supergravity and derive its corresponding bosonic function $f(R)$. An
application of our supergravity model to chaotic inflation \cite{chaot}
in early universe is discussed in sec.~4. Our conclusion is sec.~5.

\section{F(R) supergravity and its auxiliary fields}

A concise and manifestly supersymmetric description of supergravity is given
by superspace \cite{sspace}. In this section we limit our presentation to a few
basic equations. We use the units $c=\hbar=1$ and $\k=M_{\rm Pl}^{-1}$ in terms
 of the (reduced) Planck mass $M_{\rm Pl}$, with the spacetime signature 
$(+,-,-,-)$.

The chiral superspace density (in the supersymmetric gauge-fixed form) is
\be \lb{den}
\ce(x,\theta) = e(x) \left[ 1 -2i\theta\s_a\bar{\j}^a(x) +
\theta^2 B(x)\right]~, \ee
where $e=\sqrt{-\det g_{\m\n}}$, $g_{\m\n}$ is a spacetime metric, 
$\j^a_{\a}=e^a_{\m}\j^{\m}_{\a}$ is a chiral gravitino, $B=S-iP$ is the 
complex scalar auxiliary field. We use the lower case middle greek letters 
$\m,\n,\ldots=0,1,2,3$ for curved spacetime vector indices, the lower case 
early latin letters $a,b,\ldots=0,1,2,3$ for flat (target) space vector 
indices, and the lower case early greek letters $\a,\b,\ldots=1,2$ for chiral
 spinor indices.

The solution of the superspace Bianchi identitiies and the constraints defining
the N=1 Poincar\'e-type minimal supergravity results in the three relevant 
superfields $\car$, $\cg_a$ and $\cw_{\a\b\g}$ (as the parts of supertorsion), 
subject to the off-shell relations \cite{sspace}
\be \lb{bi1}
 \cg_a=\bar{\cg}_a~,\qquad \cw_{\a\b\g}=\cw_{(\a\b\g)}~,\qquad
\bar{\de}_{\dt{\a}}\car=\bar{\de}_{\dt{\a}}\cw_{\a\b\g}=0~,\ee
and
\be \lb{bi2}
 \bar{\de}^{\dt{\a}}\cg_{\a\dt{\a}}=\de_{\a}\car~,\qquad
\de^{\g}\cw_{\a\b\g}=\frac{i}{2}\de\du{\a}{\dt{\a}}\cg_{\b\dt{\a}}+
\frac{i}{2}\de\du{\b}{\dt{\a}}\cg_{\a\dt{\a}}~~,\ee
where $(\de\low{\a},\bar{\de}_{\dt{\a}}.\de_{\a\dt{\a}})$ represent the $N=1$
 supercovariant derivatives in curved superspace, and bars denote complex 
conjugation.

The covariantly chiral complex scalar superfield $\car$ has the scalar 
curvature $R$ as the coefficient at its $\theta^2$ term, the real vector 
superfield $\cg_{\a\dt{\a}}$ has the traceless Ricci tensor, 
$R_{\m\n}+R_{\n\m}-\frac{1}{2}g_{\m\n}R$, as the coefficient at its 
$\theta\s^a\bar{\theta}$ term, whereas the covariantly chiral, complex, 
totally symmetric, fermionic superfield $\cw_{\a\b\g}$ has the Weyl tensor 
$W_{\a\b\g\d}$ as the coefficient at its linear $\theta^{\d}$-dependent term. 

As regards a {\it large-scale} evolution of the FRLW Universe in terms of its 
scale factor, it is the {\it scalar} (super)curvature dependence of the 
gravitational  action that plays the most relevant role. The chiral $F(R)$ 
supergravity action, proposed in ref.~\cite{our}, reads
\be
\lb{action}
 S_{\rm sF} = \int d^4xd^2\theta\,\ce F(\car) + {\rm H.c.}
\ee
in terms of a holomorphic function $F(\car)$ of the scalar curvature superfield
$\car$. Besides manifest local $N=1$ supersymmetry, the action (\ref{action}) 
also possess the auxiliary freedom \cite{gat}, since the auxiliary field $B$ 
does not propagate. In addition, the action (\ref{action}) gives rise to the
spacetime torsion fueled by the gravitino field.

A bosonic $f(R)$ gravity action is given by 
\be \lb{mgrav}
 S_{\rm f} = \int d^4x \,\sqrt{-g}\, f(R) \ee

In order to establish a connection between the master chiral superfield 
function $F(\car)$ in eq.~(\ref{action}) and the corresponding bosonic function
 $f(R)$ in eq.~(\ref{mgrav}), we use the chiral density integration formula in
superspace $(\j_{\m}=0)$,
\be \lb{chiden}
  \int d^4xd^2\theta\,\ce \Lag =\int d^4x\, e\left\{ 
\Lag_{\rm last} +B\Lag_{\rm first}\right\} \ee 
where we have introduced the field components of the covariantly chiral 
superfield Lagrangian $\Lag(x,\theta)$, $ \bar{\de}^{\dt{\a}}\Lag=0$,  as
follows (the vertical bars denote the leading component of a superfield):
\be \lb{comps}
 \left. \Lag\right| =\Lag_{\rm first}(x)~,\qquad
 \left.\de^2\Lag\right|=\Lag_{\rm last}(x)~. \ee

In particular, we have
\be \lb{part}
 \left.\car\right| =\fracmm{\k}{3}\bar{B}=\fracmm{\k}{3}(S+iP)~,\qquad
\left.\de^2\car\right| = \fracmm{1}{3}\left( R 
-\frac{i}{2}\ve^{abcd}R_{abcd}\right) +\fracmm{4\k^2}{9}\bar{B}B~,\ee
The term $\frac{i}{2}\ve^{abcd}R_{abcd}$ does not vanish in supergravity
because of the gravitino-indiced torsion.

Appplying the chiral density formula (\ref{chiden}) to our eq.~(\ref{action}) 
yields the purely bosonic Lagrangian in the form
\be \lb{expa}
L_{\rm bos}=
F'(\bar{X}) \left[ \frac{1}{3}R_* +4\bar{X}X \right] +3X F(\bar{X})+{\rm 
H.c.}  \ee
where the primes denote differentiation. We have also introduced the notation
\be \lb{not1} X=\fracmm{\k}{3}B \qquad {\rm and} \qquad 
R_*=R-\frac{i}{2}\ve^{abcd}R_{abcd}~.\ee

Varying eq.~(\ref{expa}) with respect to the complex auxiliary fields $X$ and 
$\bar{X}$ gives rise to the algebraic equations on the auxiliary fields,
\be\lb{aux1}
3\bar{F}+X(4\bar{F}'+7F')+4\bar{X}XF'' +\frac{1}{3}F''R_*=0
\ee
and its conjugate
\be \lb{aux2}
3F+\bar{X}(4F'+7\bar{F}')+4\bar{X}X\bar{F}'' +\frac{1}{3}\bar{F}''\bar{R}_*=0
\ee
where $F=F(X)$ and $\bar{F}=\bar{F}(\bar{X})$. The algebraic equations 
(\ref{aux1}) and (\ref{aux2}) cannot be explicitly solved for $X$ in a generic 
$F(\car)$ supergravity.

\section{Our model}

Let's consider the simplest non-trivial {\it Ansatz} for the $F(R)$ 
supergravity function as
\be \lb{ansatz}
 F(\car) = -\fracmm{1}{2}f_1 \car + \fracmm{1}{2}f_2 \car^2 
\ee
with some real constants $f_1$ and $f_2$, where the first term is supposed to 
represent the standard (pure) $N=1$ Poincar\'e supergravity and the second 
term is a `quantum correction'. As regards mass dimensions of the various
quantities introduced, we have
\be \lb{dim}
[F]=3~,\quad [f_1]=2~, \quad [R]=2~, \quad [f_2]=1~,\quad [ \car ]  =1 
\ee

Being interested in the bosonic action that follows from eqs.~(\ref{action}) 
and (\ref{ansatz}), we set gravitino to zero, $\j_{\m}=0$, which also implies
$R_*=R$ and a {\it real} $X$. Equation (\ref{expa}) is now greatly simplified 
to
\be \lb{simple}
 L_{\rm bos} = 11f_2X^3 -7f_1X^2 +\fracmm{2}{3}f_2RX -\fracmm{1}{3}f_1R
\ee
 
In the limit of $f_2\to 0$ we thus have $X=0$, as it should. Hence, we recover
the Einstein-Hilbert Lagrangian
\be \lb{eh}
 L_{\rm EH} = -\fracmm{1}{3}f_1R = -\fracmm{1}{2\k^2}R= 
-\fracmm{M^2_{\rm Pl}}{2} R
\ee
provided that
\be \lb{sol1}
f_1= \fracmm{3}{2}M^2_{\rm Pl}
\ee
For a later use, we trade the parameter $f_2$ for a mass parameter $m$ as
\be \lb{set}
f_2 = \fracmm{M^2_{\rm Pl}}{m}
\ee
where $m$ is the new scale introduced in eq.~(\ref{ansatz}) 
(in addition to $M_{\rm Pl}$).

The algebraic field equation (\ref{aux1}) in our case (\ref{ansatz}) takes the
form of a quadratic equation,
\be \lb{quad}
  11X^2 -7mX +\fracmm{2}{9}R=0  
\ee
whose solution is given by
\be \lb{auxs}
\eqalign{ 
X_{\pm} &~ =~ \fracmm{7m}{22} \left[
1\pm \sqrt{1- \fracmm{8\cdot 11R}{3^2\cdot 7^2 m^2 }} \; \right]
= \fracmm{7m}{22} \left[
1\pm \sqrt{1- \fracmm{R}{R_{\rm max} }}\;  \right]  \cr 
 &~ =~  \left( \fracmm{2R_{\rm max}}{99} \right)^{1/2} \left[
1\pm \sqrt{1- \fracmm{R}{R_{\rm max} }} \; \right]  \cr }
\ee
where we have introduced the {\it maximal\/} scalar curvature
\be \lb{max}
 R_{\rm max} = \fracmm{99}{2} \left[ \fracmm{7m}{22} \right]^2
\ee
The surprising existence of the built-in maximal scalar curvature is a nice 
bonus of our construction. It comes for free, and it is very welcome for 
screening our theory of inflation from the Big Bang singularity of General 
Relativity, since eq.~(\ref{auxs}) implies $R \leq R_{\rm max}~$. This striking
property is similar to the factor $\sqrt{1-v^2/c^2}$ of Special Relativity. 
Yet another close analogy comes from the Born-Infeld non-linear extension of 
Maxwell electrodynamics, whose (dual) 
Hamiltonian is proportional to $\left( 1-\sqrt {1- \vec{E}^2/E_{\rm max}^2
-\vec{H}^2/H^2_{\rm max} +(\vec{E}\times \vec{H})^2/E_{\rm max}^2H_{\rm max}^2}
 \, \right)$ in terms of the electric and magnetic
fields $\vec{E}$ and $\vec{H}$, respectively, with their maximal values 
(see eg., ref.~\cite{bi}) for details). For instance, in string theory, one 
has $E_{\rm max}= H_{\rm max}=(2\p\a')^{-1}$.

As is clear from eq.~(\ref{auxs}), the upper bound on the scalar curvature 
exists only for $R>0$ and, in particular, for the AdS-spacetimes (in our 
notation). Equation (\ref{auxs}) does not imply an upper limit on $\abs{R}$ 
for $R<0$, in particular, for the dS-spacetimes.

Equation (\ref{quad}) can be used to reduce the Lagrangian  (\ref{simple}) to
a linear function of $X$ by double iteration. Then a substitution of the
solution (\ref{auxs}) into the Lagrangian gives us a bosonic $f(R)$ gravity 
Lagrangian  (\ref{mgrav}) in the form
\be \lb{fgr}
f_{\pm}(R) = \fracmm{-5\cdot 17 M^2_{\rm Pl} }{2\cdot 3^2\cdot 11} R
+ \fracmm{2\cdot 7}{3^2\cdot 11}M^2_{\rm Pl} 
\left(R - R_{\rm max} \right)\left[ 1\pm \sqrt{1-R/R_{\rm max} } \; \right] 
\ee

By construction, in the limit $m\to +\infty$ (or $R_{\rm max}\to +\infty$) 
both functions $f_{\pm}$ reproduce General Relativity. In another limit 
$R\to 0$, we find a cosmological constant,
\be \lb{cc}
f_-(0) \equiv \L_- =0~, \qquad f_+(0) \equiv \L_+ 
=-\,\fracmm{7^3}{2^2\cdot 11^2} M^2_{\rm Pl}m^2 =
-\, \fracmm{14}{99}M^2_{\rm Pl}R_{\rm max}
\ee
To the end of this Letter we would like to concentrate on the first solution
with the vanishing cosmological constant, so in what follows we identify 
$f_-(R)=f(R)$.

\section{$(\car +\car^2)$ supergravity model vs. $(R+R^2)$ model of inflation}

Any $f(R)$ gravity (\ref{mgrav}) is  known to be equivalent to the 
scalar-tensor gravity 
\be \lb{st}
S[g_{\m\n},\f] =  \int d^4x\, \sqrt{-g}\left\{ \fracmm{-R}{2\k^2}
+\fracmm{1}{2}g^{\m\n}\pa_{\m}\f\pa_{\n}\f - V(\f) \right\} \ee
where we have introduced the scalar (inflaton) field $\f(x)$  with its scalar
potential $V(\f)$.  The equivalence is established via a Legendre-Weyl 
transform \cite{eq}. In our notation we have~\footnote{See ref.~\cite{kkw1} for
more details. Compared to ref.~\cite{kkw1}, we changed here our notation 
$y\to -y$.}
\be \lb{nota} 
f(R)=Re^y-Z(e^y)~,\quad R=Z'(e^y)~,\quad f'(R)=e^y~,\qquad 
y= \sqrt{\fracmm{2}{3}} \fracmm{\f}{M_{\rm Pl}}
\ee
so that the inflaton scalar potential is given by \cite{kkw1}
\be \lb{spo}
 V(y) = -\fracmm{1}{2}M^2_{\rm Pl}e^{-2y}Z(e^y)
\ee

When keeping only the leading correction (beyond Einstein-Hilbert term) in
eq.~(\ref{fgr}), we get the low-curvature Lagrangian 
$(\abs{R/R_{\rm max}}\ll 1)$ in the form 
\be \lb{eff}
 f(R) = -\fracmm{1}{2}M^2_{\rm Pl}R +\a R^2 \equiv 
-\fracmm{1}{2}M^2_{\rm Pl} \left( R-R^2/M^2\right)
\ee
where
\be \lb{iden}
 \a =  \fracmm{1}{2} \fracmm{M^2_{\rm Pl}}{M^2}=
\fracmm{2}{3^3\cdot 7}\fracmm{M^2_{\rm Pl}}{m^2}
\ee
It is known as the Starobinsky model of chaotic inflation \cite{star}. The 
corresponding (inflaton) scalar potential (\ref{spo}) is well-defined and is 
given by \cite{kkw1}
\be \lb{starsp}  V(y) = V_0 \left( e^{-y}-1\right)^2 \ee
where $V_0=\fracmm{1}{8}M^2_{\rm Pl}M^2$. 
The constant term in eq.~(\ref{starsp}) 
is the vacuum energy that drives inflaton towards the minimum of the scalar 
potential (so that the inflation has an end). The conditions for a slow-roll 
(chaotic) inflation in the Starobinsky model were studied a long time ago
\cite{star}. In terms of the equivalent scalar-tensor gravity (\ref{st}) with
the scalar ponential (\ref{starsp}) we find the standard slow-roll parameters 
\cite{llbook} as follows \cite{kkw1}:
\be \lb{apeps}
\ve = \fracmm{1}{2} M^2_{\rm Pl} \left( \fracmm{V'}{V}\right)^2
= \fracmm{4e^{-2y}}{3\left( e^{-y}-1 \right)^2} =
\fracmm{3}{4N^2_e} +{\cal O}\left( \fracmm{\ln^2 N_e}{N^3_e}\right) 
\ee
and
\be \lb{apeta}
\eta =  M^2_{\rm Pl} \fracmm{V''}{V}  = 
\fracmm{4e^{-y}(2e^{-y}-1)}{3\left(e^{-y}-1 \right)^2} =
 - \fracmm{1}{N_e} + \fracmm{3\ln N_e}{4N_e^2} +\fracmm{5}{4N^2_e}
+{\cal O}\left( \fracmm{\ln^2 N_e}{N^3_e}\right) 
\ee
where the primes denote the derivatives with respect to the inflaton field 
$\f$, and the e-foldings number $N_e$ is defined by \cite{llbook} 
\be \lb{efol}
N_e = \int^{t_{\rm end}}_t H dt \approx 
\fracmm{1}{M^2_{\rm Pl}} \int^{\f}_{\f_{\rm end}} \fracmm{V}{V'} d\f
\approx \fracmm{3}{4}\left( e^y-y\right)-1.04
\ee

According to the CMB observations, the primordial spectrum in the power-law 
approximation takes the form of $k^{n_s-1}$ in terms of the comoving wave 
number $k$ and the spectral index $n_s$. For instance, the recent WMAP5 data 
\cite{wmap5} yields
\be \lb{sind5} 
 n_s = 0.960 \pm 0.013 \qquad {\rm and}\qquad r < 0.22 \ee
where $r$ is the scalar-to-tensor ratio. On the theoretical side, one has
\cite{llbook}  
\be \lb{sind}
n_s  = 1+2\h -6\ve \quad {\rm and} \quad r=16\ve
\ee
In our case, eqs.~(\ref{apeps}), (\ref{apeta}) and (\ref{sind}) imply  
\cite{kkw1}
\be \lb{apns}
 n_s =1 - \fracmm{2}{N_e} + \fracmm{3\ln N_e}{2N_e^2} -\fracmm{2}{N^2_e}
+{\cal O}\left( \fracmm{\ln^2 N_e}{N^3_e}\right) 
\ee
and
\be \lb{rho}
r = \fracmm{12}{N^2_e} +{\cal O}\left( \fracmm{\ln^2 N_e}{N^3_e}\right) 
\ee
whose leading terms agree with the earlier estimates  \cite{mchi}. It also
agrees with the WMAP5 observations (\ref{sind5}) provided that $N_e$ lies
between 36 and 71, with the average value $\bar{N}_e=54$.

The amplitude of the initial perturbations, $\D^2_R=M^4_{\rm Pl}V/(24\p^2\ve)$,
is yet another physical  observable, whose experimental value is given by 
\cite{llbook} 
\be \lb{ampl}
\left(\fracmm{V}{\ve}\right)^{1/4} =0.027\,M_{\rm Pl}
\ee
Then eq.~(\ref{ampl}) determines the normalization of the $R^2$-term in 
eq.~(\ref{action}), in agreement with earlier calculations (see eg., 
ref.~\cite{rus}),
\be \lb{scale} \fracmm{M}{M_{\rm Pl}}= (3.5\pm 1.2)\cdot 10^{-5}
\ee
where we have used $N_e=\bar{N}_e=54$. In the case (\ref{iden}) we find
\be \lb{scalem}
m =\fracmm{2}{3\sqrt{21}} M\approx 0.15\, M 
\approx 5\cdot 10^{-6}\,M_{\rm Pl} \quad {\rm and} \quad
R_{\rm max} \approx 10^{-10}\,M^2_{\rm Pl} 
\ee
Unfortunately, it also implies that we {\it cannot} embed the Starobinsky
$(R+R^2)$-type inflationary model into our $(\car+\car^2)$ supergravity, 
because the higher-order curvature terms cannot be ignored in eq.~(\ref{fgr}),
ie. the $R^n$-terms with $n\geq 3$ are not small against the $R^2$-terms, and
$\abs{R/R_{\rm max}}\sim {\cal O}(1)$ during inflation. For example, in the
expansion
\be \lb{cube}
f_-(R)= -\fracmm{1}{2}M^2_{\rm Pl}\left(  R - \fracmm{R^2}{M^2}
 - \fracmm{R^3}{7M^4}\right)  +{\cal O}(R^4)
\ee
the $R^3$-term is already not negligible.

The {\it exact} gravitational function $f_-(R)$ in eq.~(\ref{fgr}) also leads 
to a well-defined (single-valued, non-singular, bounded from below) inflaton 
scalar potential,
\be \lb{expot}
V(y) = V_0 \left( 11e^y +3\right)\left(e^{-y}-1\right)^2
\ee
where $V_0= (3^3/2^6)M^2_{\rm Pl}m^2$. The associated slow-roll inflation
parameters are given by
\be \lb{exeps}
\ve (y) = \fracmm{1}{3} \left[ \fracmm{ e^y\left( 11+ 11e^{-y} +6 e^{-2y}
\right) }{ (11e^y+3)
(e^{-y}-1)} \right]^2  \geq \fracmm{1}{3} \ee
and
\be \lb{exeta}
\eta(y)= \fracmm{2}{3} \fracmm{\left(11 e^y+5e^{-y}+12 e^{-2y}\right) }{ 
(11e^y +3) (e^{-y}-1)^2 } \geq \fracmm{2}{3}
\ee
which are not small enough for matching the observational data (WMAP). Unlike
the potential (\ref{starsp}), the potential (\ref{expot}) is too steep to 
support a slow-roll inflation. A possibility of destabilizing the 
Starobinsky cosmological scenario (based on adding the $R^2$-term to the 
Einstein-Gilbert term) against the terms of the higher order with respect to 
the scalar curvature was observed earlier in ref.~\cite{mae}.

\section{Conclusion}

Our main new result is given by eq.~(\ref{fgr}). The $f(R)$ gravity with that 
function can be locally $N=1$ supersymmetrized to the $F(R)$ supergravity
described by eq.~(\ref{ansatz}). That $F(R)$ supergravity model has the upper 
bound on the scalar curvature --- see eq.~(\ref{max}). The possible existence 
of such bound in supergravity was conjectured in ref.~\cite{gk}. 

Unfortunately, the function (\ref{fgr}) is not suitable for a slow-roll 
inflation. It is worth noticing that it does not mean the failure of the whole 
approach ($F(\car)$ supergravity), because the function (\ref{ansatz}) was
chosen {\it ad hoc}, due to its simplicity only. It is conceivable that 
there exist many other functions $F(\car)$ leading to slow-roll inflation in 
agreement with the observations. 

Revealing the (quantum) origin of the higher-order scalar supercurvature terms 
in the supergravity function $F(\car)$ is beyond the scope of this paper. For
instance, they may come through the radiative corrections responsible for the
anomalies of some classical symmetries (like K\"ahler symmetry) in the 
matter-coupled supergravities \cite{ovr}, or they may come from Superstrings 
\cite{our}.  

\section*{Acknowledgements}

The author is grateful to the Theory Division of CERN for kind hospitality
extended to him during preparation of this paper. He also thanks George Dvali
and Alexei Starobinsky for discussions and correspondence.

\end{document}